\documentclass[pre,twocolumn,groupedaddress,showpacs,showkeys,amsmath]{revtex4}

\usepackage{graphicx}
\usepackage{dcolumn}
\usepackage{bm}

\begin{document}


\title{Analysis of the phase transition in the $2D$ Ising ferromagnet using a Lempel-Ziv string parsing scheme and black-box data-compression utilities}

\author{O. Melchert$^1$}
\email{oliver.melchert@uni-oldenburg.de}
\author{A. K. Hartmann$^1$}
\email{alexander.hartmann@uni-oldenburg.de}
\affiliation{
$^1$ Institut f\"ur Physik, Universit\"at Oldenburg, Carl-von-Ossietzky Strasse, 26111 Oldenburg, Germany\\
}

\date{\today}


\begin{abstract}
In this work we consider information-theoretical observables to
analyze short symbolic sequences, comprising time-series that represent
the orientation of a single spin in a $2D$ Ising ferromagnet on a square lattice 
of size $L^2=128^2$, for different
system temperatures $T$. The latter were chosen from an interval enclosing 
the critical point $T_{\rm c}$ of the model. At small temperatures the sequences
are thus very regular, at high temperatures they are maximally random. In the
vicinity of the critical point, nontrivial, long-range correlations appear.
Here, we implement estimators for the entropy rate, excess entropy (i.e.\
``complexity'') and multi-information.
First, we implement a Lempel-Ziv string parsing scheme, 
providing seemingly elaborate entropy rate
and multi-information estimates and an approximate estimator for the excess
entropy. Furthermore, we
apply  easy-to-use black-box data compression utilities, providing
approximate estimators only. For comparison and
to yield results for benchmarking purposes we implement the 
information-theoretic observables also
 based on the well-established $M$-block Shannon entropy, which is
more tedious to apply compared to the 
the first two ``algorithmic''
entropy estimation procedures. To test how well one can 
exploit the potential of such data
compression techniques, we aim at detecting the critical point the $2D$ Ising ferromagnet.
  Among the above observables, the multi-information, which
is known to exhibit an isolated peak at the critical point, is very easy to
replicate by means of both efficient
 algorithmic entropy estimation procedures.
Finally, we assess how good the various algorithmic entropy estimates compare to 
the more conventional block entropy estimates and illustrate a 
simple modification that yields enhanced results. 
\end{abstract} 

\maketitle

\section{Introduction}
\label{sect:introduction}

The standard analysis of phase transitions in terms of statistical mechanics
involves the analysis of order parameters and other derivatives of the free
energy, related to a given model system \cite{goldenfeld1992}. Routinely one
studies model systems that involve many degrees of freedom and local
interactions that nevertheless result in a non-trivial, seemingly ''complex''
behavior.  Implying a rather naive use of the word, such systems are regarded
as being very ''complex'', particularly right at the point where a phase
transitions occurs in the underlying model.  From a point of view of
statistical mechanics, a large degree of complexity is shown by growing
correlations as one approaches the critical point by tuning a proper system
parameter.  Correspondingly, throughout the analysis of observables related to
such model systems it is often desirable to find a measure for what is naively
referred to as the ''complexity'' of the underlying system \cite{Grassberger1986,Feldman1998}. However, a precise
definition of the term ''complexity'' is often elusive.  An alternative
approach to the analysis of phase transitions, which has recently gained
popularity in the analysis of complex systems, is based on a purely information
theoretic approach \cite{Feldman2008,Crutchfield2003,crutchfield2010}.  A variety of previous studies
employed such information-theoretic methods to measure the entropy rate (i.e.\ disorder
and randomness) and statistical complexity (i.e.\ structure, patterns and
correlations) for $d\geq1$-dimensional systems
\cite{Arnold1996,Crutchfield1997,Feldman2003,Feldman2008,Robinson2011,Wilms2011,Melchert2013}.  In particular for
$1D$ systems, the excess entropy constitutes a well understood
information-theoretic measure of complexity, providing a well defined literal
sense for that term.  Effectively, the excess entropy accounts for the rapidity
of entropy convergence. 
In order to obtain numerical values for the entropy rate and
complexity, the well 
established information-theoretic approach presented in Ref.\ \cite{Crutchfield2003} is based on 
the notion of a ``block entropy'', see discussion below.  Among several
intriguing findings, it led to the analysis of
complexity-entropy diagrams that allow for a characterization of the temporal
and spatial dynamics of various stochastic processes, including simple maps as
well as Ising spin-systems, in purely information-theoretic coordinates
\cite{Feldman2008}. 

On the other hand, note that there are a variety of other measures for what is
known as algorithmic entropy, as, e.g., the description size of a minimal algorithm (or computer,
or circuit) which is able to generate an instance of the problem under scrutiny
\cite{kolmogorov1963,chaitin1987,machta2006}.  However, such measures are often
impractical when it comes to the analysis of large systems. 
In this regard, as discussed in the literature \cite{Ebeling1997,Puglisi2003}, 
particular data compression algorithms might render a natural and particularly
simple candidate for estimating an algorithmic entropy.
The pivotal challenge of such data compression algorithms, readily available as
black-box data compression utilities as, e.g., the {\tt zlib} \cite{zlib_ref}, 
{\tt bzip2} \cite{bzip2_ref} and {\tt lzma} \cite{lzma_ref} utilities, is to discover
\emph{patterns} (synonymous with regularities, correlations, symmetries and
structure; see Sect.\ II of Ref.\ \cite{Shalizi2001}) in the given 
input data and to exploit the respective redundancies in order to minimize the
space required to store the data.  Interestingly, the pattern discovery and
data compression process of particular data compression schemes finds
application in contexts as diverse as, e.g., DNA sequence classification
\cite{Loewenstern1995}, entropy estimation \cite{Ebeling1997,Baronchelli2005,Grassberger2002}, 
and, more generally, time series analysis \cite{Puglisi2003}.
However, at this point please note that not all of the applications
of such methods reported in the scientific literature are without critique,
see \cite{Benedetto2002,Khmelev2003,Goodman2002}.

In the presented study we aim to assess how well algorithmic entropy (AE)
estimates, obtained using a Lempel-Ziv (LZ) string parsing scheme \cite{Lesne2009,Liu2012} and black-box data compression utilities,  and the
results obtained therewith compare to those obtained by means of the respective
block entropy (BE) estimates used in the context of the information theoretic
approach mentioned earlier.  
As raw data, to be processed further by both
approaches, we consider binary sequences that represent the spin-flip dynamics,
induced by single-spin-flip Metropolis updates \cite{newman1999} of the $2D$ Ising Ferromagnet
(FM) on a square lattice of side length $L=128$ with
fully periodic boundary conditions at different temperatures $T$. The temperatures are chosen from the interval 
$T\in[2,\,2.8]$, enclosing the critical point $T_{\rm c}=2.269\ldots$ of the model. 
In order to model the binary sequences, a particular spin on the 
lattice is chosen as a ``source'', emitting symbols from the binary alphabet
$\mathcal{A}=\{0,1\}$ (after a simple transformation of the spin variables). 
Therefore, the orientation of the source-spin is
monitored during a number of $N$ Monte Carlo (MC) sweeps to yield symbolic sequences 
 $S=(s_1,\ldots,s_N)$ of length up to $N=5 \times 10^5$. 
Before the spin orientation is recorded, a sufficient number 
of sweeps are performed to ensure that the system is equilibrated.
In this regard, for a square lattice with side length $L=128$, 
and starting with all spins ``up'', and by analyzing the magnetization
of the system, we observed an equilibration time of approximately 
$\tau_{\rm eq}=3000$ MC 
sweeps for the lowest temperature. 
However, for each system considered we discarded the first $10^5$ sweeps to avoid
initial transients.

The aim of this work is use computer science methods, related to the field of
lossless data compression, and apply them to a pivotal model system from
statistical mechanics, namely the $2D$ Ising ferromagnet and the continuous
ferromagnet-to-paramagnet transition found therein. In particular, we want to
clarify whether the phase transition can be detected, located and analyzed
numerically with high precision just by looking at entropy and complexity
measures derived via data compression utilities. Being well aware that such AE
estimates based on sequence parsing schemes and data compression utilities might only be used to obtain
upper bounds on the actual entropies of the underlying (finite) symbolic
sequences \cite{Ebeling1997,Schuermann1996,Lesne2009}, we compare our findings to those obtained using the BE estimators that
here serve as a benchmark.

The remainder of the presented article is organized as follows.
In section \ref{sect:infTheor} we introduce the information-theoretic
observables obtained from the limiting behavior of block entropies and we detail the LZ string parsing and
data compression based entropy measures. In section \ref{sect:results} we discuss
the results obtained by applying the aforementioned entropy estimators and 
in section \ref{sect:summary} we conclude with a summary. 

\section{Information-theoretic observables for symbolic sequences}
\label{sect:infTheor}

In subsection \ref{ssect:infTheor_1} we introduce the basic notation from information theory,
subsequently used to define the entropy rate, excess entropy and further related measures that might
be associated to a one-dimensional ($1D$) symbolic sequence of finite length. 
Regarding the definition of the entropy rate and excess entropy we 
follow the notation used in Refs.\ \cite{Shalizi2001,Crutchfield2003,Feldman2008}, 
where a more elaborate discussion of the individual information-theoretic 
observables can be found.
In subsection \ref{ssect:infTheor_2} we further introduce the LZ string parsing scheme and 
data compression based entropy measures considered in the remainder.

\subsection{Block entropy, entropy rate and complexity}
\label{ssect:infTheor_1}

Given a symbolic sequence $S$ of finite length $N$, i.e.\
$S=(s_1,s_2,\ldots,s_N)$, where the individual symbols $s_i$ assume a symbolic
value randomly drawn from an alphabet $\mathcal{A}$ of finite size.  Here, an individual
symbol signifies the outcome of a measurement on a random variable, i.e.\ the
orientation of a single Ising spin at a given point in (simulation) time.
Therefore, unless otherwise specified, $\mathcal{A}$ will denote the binary
alphabet $\{-1,+1\}$.  For $s^M$ denoting a particular symbol block of length
$M>0$, the $M$-block Shannon entropy, also called ``block entropy'' (BE), a prerequisite needed to define the
subsequent information theoretic observable, reads 
\begin{align} 
H_{{\rm BE},M}[S]\equiv-\sum_{s^M\in\mathcal{A}^M} {\rm Pr}(s^M) \log_2({\rm Pr}(s^M)) \label{eq:blockEntropy},
\end{align} 
wherein ${\rm Pr}(s^M)$ specifies the joint probability for blocks of $M$
consecutive symbols. Hence, considering a finite sequence,  ${\rm Pr}(s^M)$
represents the empirical rate of occurrence of $s^M$ in the given sequence.
Consequently, $H_{{\rm BE},N}$ depends on the spin-flip dynamics of the chosen Ising spin,
implemented by the simulation procedure for the $2D$ Ising FM, over intervals
of $M$ consecutive time steps.
In the above formula, the sum runs over all possible $M$-blocks, i.e., combinations
of $M$ consecutive symbols, that might be composed by means of the alphabet
$\mathcal{A}$ (considering a binary alphabet, there are $2^M$ such blocks) and we
imply to set $0 \log_2(0) \equiv 0$.
In general, $H_{{\rm BE},M}$ is a nondecreasing function of $M$, bounded 
by $H_{{\rm BE},M}\leq M\cdot \log_2{|\mathcal{A}|}$. The upper bound is attained in the 
limit of sequence length $N\to\infty$ if 
the probability of a string factorizes and each letter has the
same probability of occurrence, i.e., ${\rm Pr}(s^M)=1/|\mathcal{A}|^M$.
In the limit of large block-sizes and sequence lengths $N$, $H_{{\rm BE},M}$ might thus not converge to a finite
value. As a remedy, due to the above bounding value, the \emph{entropy rate} 
\begin{eqnarray}
h \equiv \lim_{M,N\to \infty} H_{{\rm BE},M}[S]/M. \label{eq:entropyRate}
\end{eqnarray}
might be considered instead. It specifies the asymptotic rate of increase of
the $M$-block Shannon entropy with block length $M$, and, in the limit of large
block size and sequence length, indicates an upper bound on the number of bits
needed to encode a symbol of the observed sequence.

In most practical applications the finite length of the underlying symbolic
sequences imposes certain sampling issues related to subsequences of a long
enough length $M$, rendering it unfeasible to proceed towards very large block
sizes. E.g., for sequences of length $N$, where symbols are independent and
identically distributed (iid), one might experience a severe undersampling of
$M$-blocks if, at a given alphabet size $|\mathcal{A}|$, $M$ is too large or
$N$ is too short. In particular, a naive upper bound $M_{\rm max}$ might be
obtained from the constraint $N\geq M 2^M$ \cite{Lesne2009,Schuermann1996}.
Consequently, it is desirable to consider proper finite-$M$ approximations to
the entropy rate observed for sequences of finite length $N$, referred to as \emph{apparent entropy rates}.
Two such estimators are given by the \emph{per symbol} $M$-block entropy
\begin{eqnarray}
h^{\prime}_{{\rm BE},M}[S] = H_{{\rm BE},M}[S]/M  \label{eq:entropyRate_perSymbol}
\end{eqnarray}
and the discrete derivative of Eq.\ \ref{eq:entropyRate}, defining the
entropy rate as a block entropy increment via
\begin{eqnarray}
h_{{\rm BE},M}[S] = H_{{\rm BE},M}[S]-H_{{\rm BE}, M-1}[S], \label{eq:entropyRate_difference}
\end{eqnarray}
for sequences of finite length $N$ and both equations presuming $M\geq 1$. Viewed as a function of block size,
the finite-$M$ estimates of the entropy rates converge to the asymptotic value
$h$ from above. Hence, considering small block sizes, the underlying symbolic
sequences tend to look more random than in the limit $M\to\infty$.  Finally,
note that the entropy rate is a measure of \emph{randomness} that might be
attributed to the underlying sequences \cite{Feldman1998,Crutchfield2003}.
Here, for sequences of finite size $N$ and for a maximally feasible block-size $M_{\rm max}$ 
we denote the block entropy based estimate of the entropy rate as
\begin{align}
h_{\rm BE}[S] \equiv h_{{\rm BE},M_{\rm max}}[S].
\end{align}

For one-dimensional symbolic sequences, there are three different but
equivalent expressions to define the excess entropy. These are based on the
convergence properties of the entropy rate, the subextensive part of the
block entropy in the limit of large block sizes and the mutual information
between two semi-infinite blocks of variables, see Refs.\
\cite{Feldman2003,Feldman2008}.  Here, we focus on the definition of the excess
entropy, also termed ``(statistical) complexity'', related to the
convergence properties of the entropy rate in the form
\begin{align}
C_{\rm BE}[S]=\sum_{M=1}^{\infty} (h_{{\rm BE},M}[S]-h_{\rm BE}[S]). \label{eq:excessEntropy}
\end{align}
As pointed out above, the conditional entropies $h_{{\rm BE},M}[S]$ constitute upper
bounds on the asymptotic entropy rate, allowing, in principle, for an improving
estimate of $h_{\rm BE}$ for increasing $M$. Note that since the sum in Eq.\ \ref{eq:excessEntropy}
extends to $M\to\infty$, it implies the limit $N\to\infty$ and $h_{\rm BE}[S]=h$. However, for practical purposes, i.e.\ since
we consider sequences of finte lenght $N$ only, the sum
in Eq.\ \ref{eq:excessEntropy} needs to be truncated at a maximally feasible
block-size $M_{\rm max}$ that still yields a reliable estimate of $h$ (see
discussion above). Hence, for a symbolic sequence $S$ of finite length $N$, $C_{\rm BE}[S]$ 
accounts for the randomness that is present at small values of $M$ that vanishes in the limit of large block sizes. The excess
entropy is considered a measure of \emph{statistical complexity}
\cite{Feldman1998,Crutchfield2003} with the ability to detect structure within
the considered sequences \cite{Grassberger1986}. Thereby it satisfies the desirable ``one-hump''
criterion \cite{Crutchfield2000}, according to which a proper measure for
statistical complexity yields a small numerical value for highly ordered and highly
disordered sequences.
Furter, note that also other
practically computable approaches exist, which indeed seem
to measure complexity as expected, like mutual information 
\cite{Grassberger1986,cover2006} or statistical complexity 
\cite{Crutchfield1989}.

For $1D$ symbolic sequences, a further information-theoretic observable, termed
\emph{multi information} \cite{Erb2004} is given by the first summand in Eq.\
\ref{eq:excessEntropy}, i.e.\
\begin{align}
I_{\rm BE}[S]=h_{{\rm BE},1}[S]-h_{\rm BE}[S]. \label{eq:multiInformation}
\end{align}
Albeit $I_{\rm BE}$ is closely related to the excess entropy $C_{\rm BE}$ (this holds only in the
limit of large block sizes $M$; see Ref.\ \cite{Erb2004} for a more general
discussion of the multi information), it captures a particular contribution to
the convergence of the entropy density.  In this regard, for sequences of infinit length, 
i.e.\ in the limit $N\to\infty$, it measures the
decrease of the entropy rate observed by switching from the level of
single variables ($1$-block) statistics to the statistics attained as
$M\to\infty$.
Recently, the multi-information was introduced and used to
characterize spin configurations for the 2D Ising FM in the thermodynamic limit
by analytic means \cite{Erb2004}. It was found to exhibit an isolated global
maximum right at the critical temperature $T_{\rm c}$ and thus also satisfies
the ``one-hump'' criterion desired for the full complexity measure.

\subsection{String parsing and data compression based entropy measures}
\label{ssect:infTheor_2}

Below we illustrate the Lempel-Ziv string parsing scheme and the data-compression tools
used to implement entropy measures by algorithmic means.

\paragraph{Lempel-Ziv string parsing scheme:}

The Lempel-Ziv (LZ) string parsing scheme considered here is based on a coarse graining of 
the input sequence $S$, yielding a quantity termed ``Lempel-Ziv'' complexity.
Therefore, the input sequence is split into several independent
blocks. By traversing the input sequence, a new block is completed whenever one
encounters a new subsequence of consecutive symbols that does not match a
subsequence of the already traversed part of the input sequence.

More formally, given the symbolic sequence $S=(s1,\ldots,s_N)$, for which the 
subsequence $(s_i,\ldots,s_j)$ might be specified as $S(i,j)$, we obtain a parsed
sequence $S^\prime$ using the following procedure:
\begin{itemize}
\item[(1)]  initialize the parsed sequence $S^\prime=(s_1)$ and an empty auxiliary sequence $Q=()$. 
Traverse the sequence $S$ from left to right, using an index $i$ initialized as $i=1$.
\item[(2.1)] In step $i$, advance $i$ to $i+1$ and extend the auxiliary sequence $Q$ via symbol $s_{i+1}$.\\ 
E.g., after increasing $i=1\to i=2$ one has $Q=(s_2)$. More general, if the auxiliary sequence
already reads $Q=(s_j,\ldots,s_i)$, it is extended to $Q=(s_j,\ldots,s_i,s_{i+1})$. 
\item[(2.2)] Check whether the current auxiliary sequence $Q$ matches any subsequence of $S(1,i)$.
If not, append the auxiliary sequence $Q$ as a new ``block'' to the parsed sequence $S^\prime$ and reset
$Q=()$. \\
E.g., in step $1$ one has $Q=(s_2)$. If $Q$ does not match 
the symbol $S(1,1)=s_1$, then
set $S^\prime=(s_1)(s_2)$ and reset $Q=()$.
\item[(3)] repeat (2.1) and (2.2) until $i=N$ and append $Q$ to $S^\prime$ to yield the final parsing $S^\prime$.  
\end{itemize}
Finally, the
LZ complexity $N_{\rm LZ}[S]$ associated to sequence $S$ is simply the number of consecutive blocks found after the coarse-graining 
procedure is completed \cite{Liu2012}. As an example, consider the sequence $S=1010000110$. Following the above procedure yields the
parsed sequences $S^\prime=(1)(0)(100)(001)(10)$ for which the LZ complexity reads $N_{\rm LZ}[S]=5$.

The LZ complexity provides means to quantify the degree of order or disorder in an observed 
symbolic sequences \cite{Liu2012,Lesne2009}. Therefore the term ``complexity'' is somewhat misleading in our context.
As pointed out in Ref.\ \cite{Lesne2009}, a proper normalization allows to relate the 
LZ complexity to the entropy rate of the symbolic sequence, i.e.\
\begin{align}
h_{\rm LZ}[S] = \lim_{N\to\infty} h_{\rm LZ}[S] = \lim_{N\to\infty} \frac{N_{\rm LZ}[S] \ln(N)}{N}. \label{eq:LZcplx_eRate}
\end{align} 
Note that in Ref.\ \cite{Lesne2009} two different variants of LZ parsing are
considered, here we use the parsing scheme that Ref.\ refers to as LZ $77$.

The above string parsing scheme might also be used to compute an observable
that closely follows the definition of the block-entropy based excess entropy.
Therefore, bear in mind that in the definition of the excess entropy Eq.\
\ref{eq:excessEntropy}, the individual terms involve the entropy-rate estimates
$h_{{\rm BE},M}$ for finite block-size $M$, i.e.\ containing symbol correlations up to length
$M$, only. By using the LZ string parsing scheme, similar estimates of the
entropy rate, restricted to feature correlations up to some specified length
$M$, might be obtained by preprocessing the initial sequence by applying a $M$-block standard random shuffle procedure.  
Thereby, the initial length-$N$ sequence $S$ is
first split into $\lfloor N/M \rfloor$ blocks of length $n$ and possibly a
remaining block of length $N {\rm mod}(M)$. Then, these blocks are brought into
random order and merged to form the new, $M$-block shuffled surrogate sequence
$S^{(M)} = {\rm shuffle}(S,M)$.  This maintains the individual symbol
frequencies, destroys all correlations that extend over lengths larger than $M$
and yields a particular realization of a $M$-block shuffled surrogate sequence.
Note however that the distribution of $M$-blocks, obtained by sliding a window
of length $M$ over the initial sequence and keeping track of all overlapping
subsequences of length $M$ (e.g.\ used to compute $M$-block entropies), is not
conserved by this procedure. The case $M=1$ corresponds to the standard random shuffling 
procedure considered in Ref.\ \cite{JimenezMontano2002b}. Finally, a standard random shuffle 
based excess entropy (or similar: standard random shuffle based complexity)
that utilizes the LZ string parsing scheme might be obtained as
\begin{align}
C_{\rm s}[S] = \sum_{M=1}^{M_{\rm max}} (h_{\rm LZ}[S^{(M)}] - h_{\rm LZ}[S]), \label{eq:randShuffleComplexity}
\end{align}
where $M_{\rm max}$ indicates a maximal feasible block-size for the shuffling procedure.
Albeit the above observable is no direct analog of the excess entropy Eq.\
\ref{eq:excessEntropy}, it is expected to behave in a similar manner.

Similarly, the string parsing scheme might be used to compute a shuffling based
equivalent of the multi-information \cite{Erb2004} Eq.\ \ref{eq:multiInformation} as
\begin{align}
I_{\rm s}[S] = h_{\rm LZ}[S^{(1)}] - h_{\rm LZ}[S]. \label{eq:randShuffleMultiInf}
\end{align}
Therein, $S^{(1)}$ simply represents a surrogate sequence obtained by a
$1$-block standard random shuffle \cite{JimenezMontano2002b} which maintains
the symbol frequencies but destroys correlations on all scales. Hence, Eq.\
\ref{eq:randShuffleMultiInf} reflects the entropy overestimate observed by
going from a representation of the sequence with no correlations at all to its
original representation including all correlations. 

\paragraph{Data-compression tools:}

In addition we also consider three commonly used black-box data compression 
tools, namely {\rm zlib} \cite{zlib_ref}, {\rm bz2} \cite{bzip2_ref}, and {\rm lzma} \cite{lzma_ref}, 
in order to compute an algorithmic entropy that is based on the compressibility of the
underlying sequence according to 
\begin{align}
h_{\rm alg}[S] = \frac{{\rm length}({\rm compress}(S))}{N}, \label{eq:algEntropy}
\end{align}
see Ref.\ \cite{Ebeling1997}. According to the latter reference, these data-compression 
tool based entropy should provide an upper bound on the entropy of the underlying sequences.
Note that the shuffling based (approximate) excess entropy and multi-information can 
also readily be computed using Eq.\ \ref{eq:algEntropy}.


\section{Results}
\label{sect:results}

Subsequently we will address two distinct issues: Firstly, in subsection
\ref{subsect:locateCritPt}, we will assess how well the aforementioned,
string-parsing and data-compression based entropy estimators might be used to
characterize the ferromagnet-to-paramagnet transition in the $2D$ Ising
ferromagnet. Therefore we consider the information theoretic observables
introduced previously in section \ref{sect:infTheor} and capitalize on their
scaling behavior as function of the system temperature. Since these estimates
are obtained by algorithmic means, we here refer to the them as ``algorithmic
entropy estimates''. Secondly, in subsection \ref{subsect:compareEntropy}, we
evaluate how well the algorithmic entropy estimates compare to the more
conventional block entropy estimates and discuss further means to improve on the
difference between the respective estimates.

\subsection{Using algorithmic entropy estimates to locate the critical point}
\label{subsect:locateCritPt}
In the presented subsection we will summarize the results on the issue of how
well the phase transition in the $2D$ Ising FM can be resolved by means of the
Lempel-Ziv (LZ) string parsing scheme \cite{Lesne2009,Schuermann1996,Liu2012},
as well as the {\rm zlib} \cite{zlib_ref},
{\rm bz2} \cite{bzip2_ref}, and {\rm lzma} \cite{lzma_ref} black-box data
compression utilities.  Therein, we are not interested in the absolute values
of the entropy estimates thus obtained, but merely in the data-curve
characteristics as function of the system temperature. The final estimates for
the transition points, resulting from the considered observables, and their
comparison to the literature value of the critical temperature, i.e.\ $T_{\rm
c}^{\rm lit}=2.269$, can be used to assess the value of the algorithmic entropy
estimators as easy-to-compute utilities, that might yield valuable information
on the structural change as visible in measurements of finite-length symbolic sequences.

\subsubsection{Lempel-Ziv string parsing scheme}

\begin{figure}[t!]
\begin{center}
\includegraphics[width=0.9\linewidth]{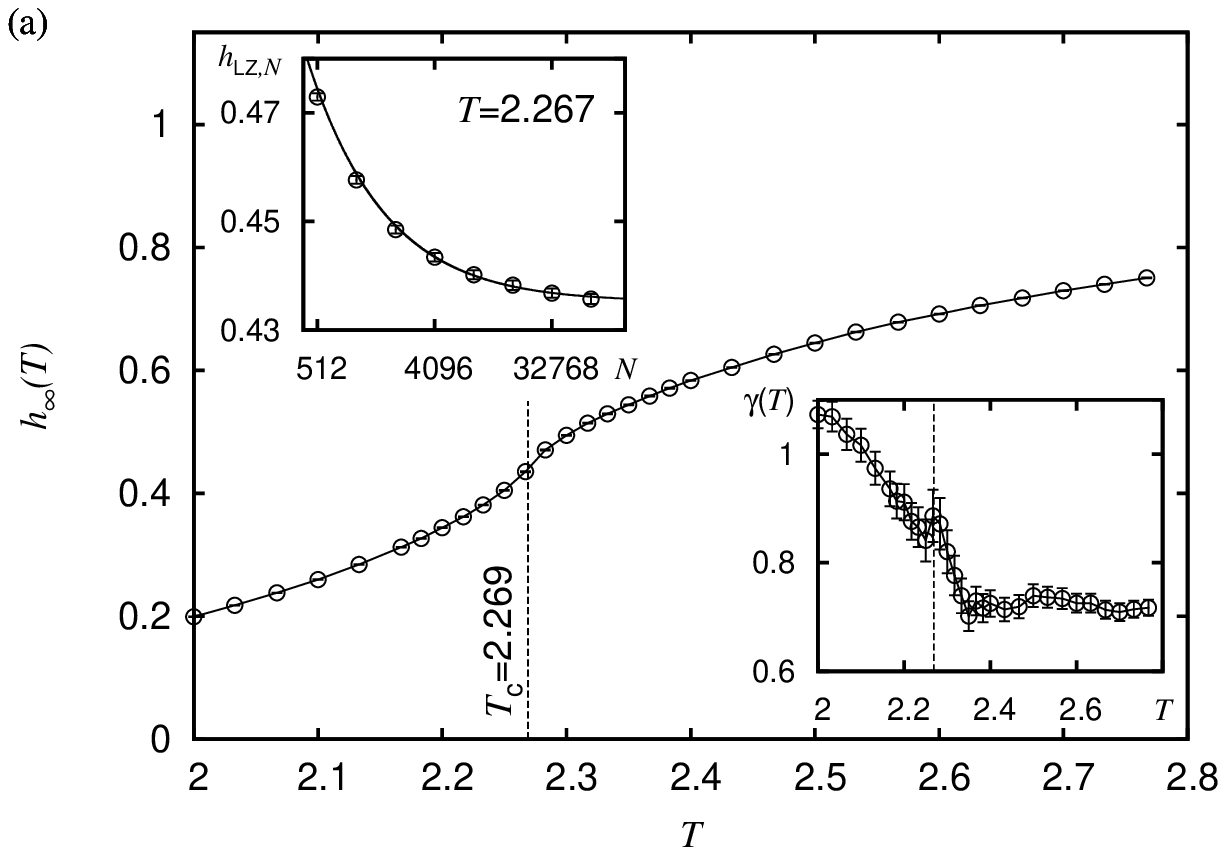}
\includegraphics[width=0.9\linewidth]{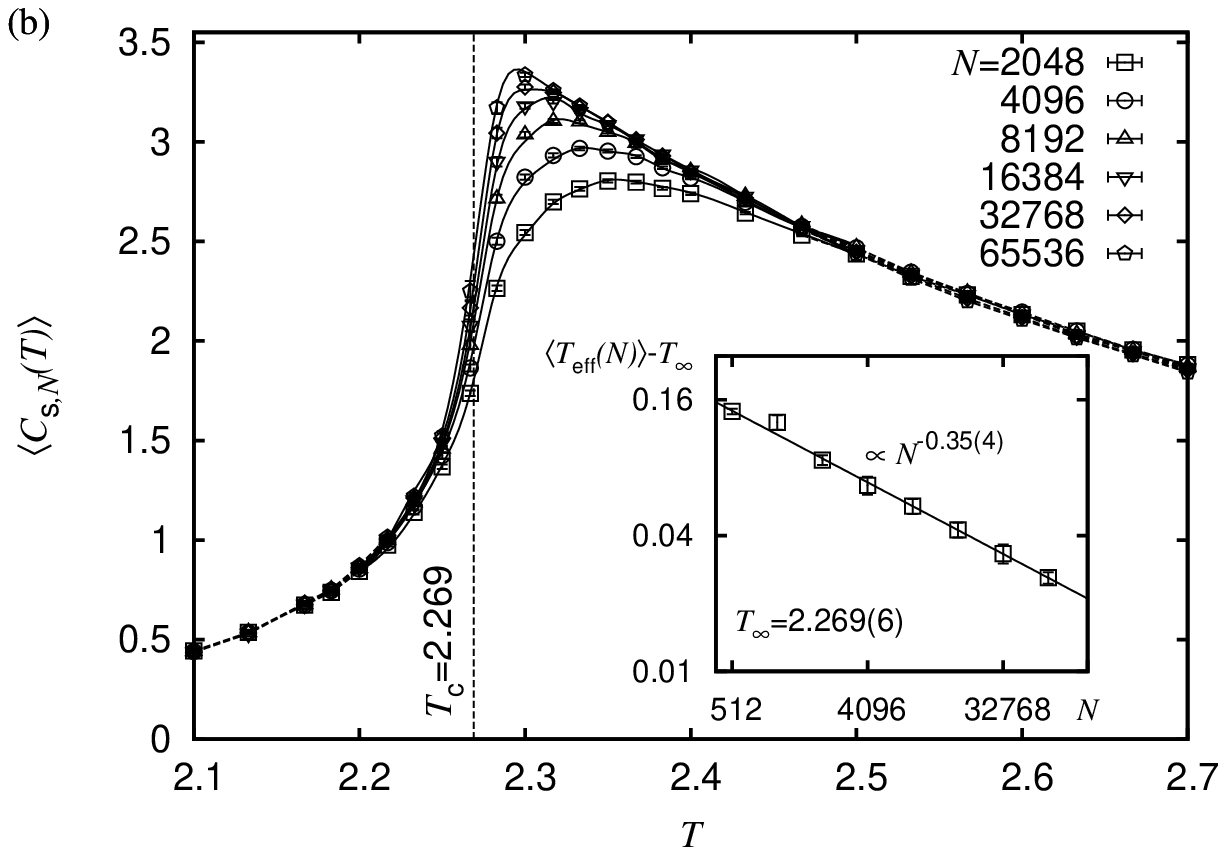}
\includegraphics[width=0.9\linewidth]{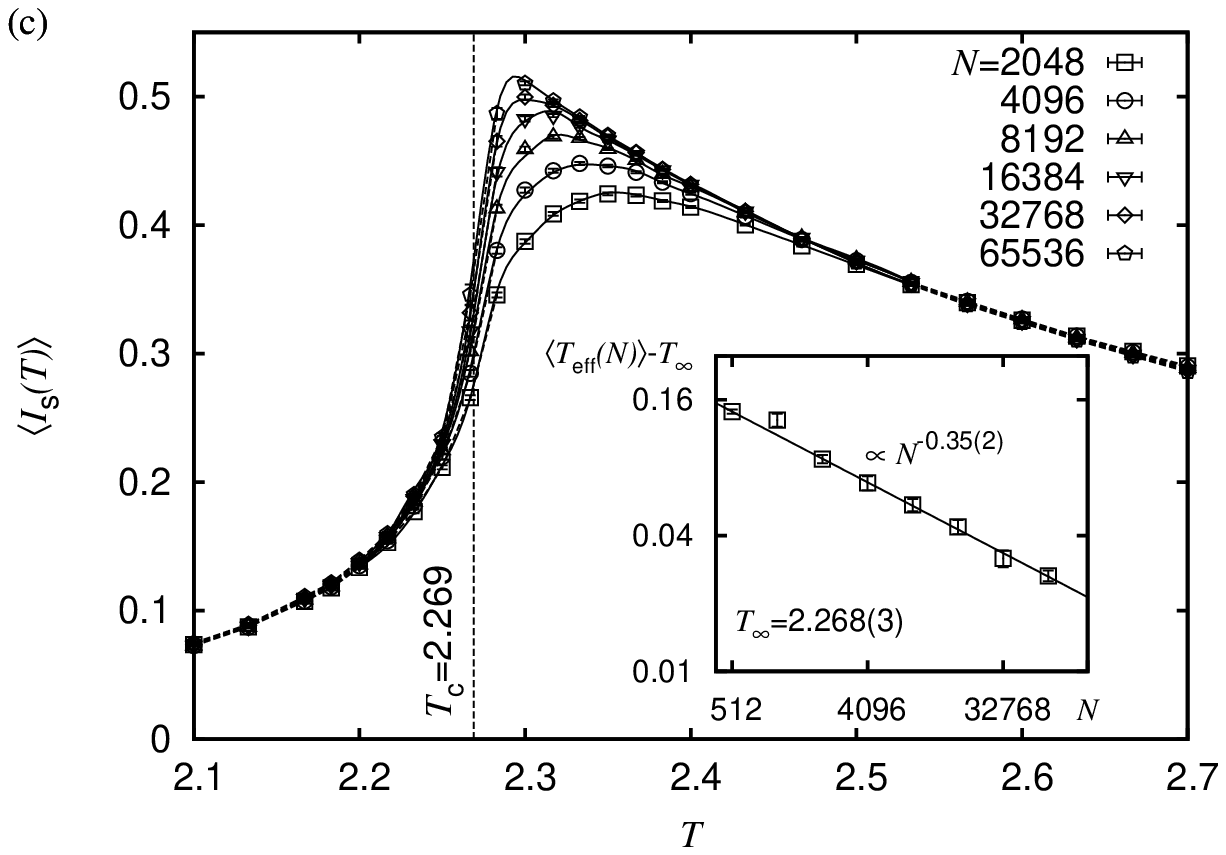}
\end{center}
\caption{\label{fig:LZcplx}
Results for the information theoretic observables computed using the Lempel-Ziv
sequence parsing scheme averaged over many input sequences.  
(a) Extrapolated estimates $h_\infty(T)$ of the entropy rate as function of the
system temperature $T$. The upper inset illustrates the extrapolation of the
finite-length entropy-rate estimates $h_{{\rm LZ},N}$ at $T=2.267\approx T_{\rm c}$ using the fit function
discussed in the text. The lower inset shows the fit-exponents $\gamma(T)$ as
function of the temperature.
(b) Analysis of the block-shuffling based excess entropy, also termed ``complexity'',
$C_{\rm s}(T)$. The inset illustrates the extrapolation of the system-size dependent peak-locations $T_{\rm eff}(N)$
to the asymptotic limit.
The solid lines indicate cubic spline, fitted to the interval $T\in[2.24,2.5]$ and the 
dashed lines are a guide to the eyes, only.
(c) Analysis of the block-shuffling based multi-information $I_{\rm s}(T)$ similar to (b).
}
\end{figure}

\paragraph{Entropy rate:}
As pointed out previously, for a given input sequence $S$, consisting of $N$
consecutive symbols, the LZ string parsing scheme yields a parsed sequence of
length $N_{\rm LZ}[S]$, allowing to compute the respective entropy rates 
for sequences of finite length via $h_{{\rm LZ},N}[S]$, see Eq.\ \ref{eq:LZcplx_eRate}.
We applied this estimation procedure to ensembles of length-$N$ sequences,
accounting for the orientation of a selected spin in a $2D$ Ising FM at different
values of the temperature parameter $T$, considering various values of $N$ up
to $N=2^{16}=65536$, see Fig.\ \ref{fig:LZcplx}(a).  Therein, the upper inset
of Fig.\ \ref{fig:LZcplx}(a) illustrates the change of the average asymptotic
entropy rate at $T=2.267\approx T_{\rm c}$ with increasing sequence length $N$.
For the extrapolation to the asymptotic limit, an empirically motivated fit
function of the form 
\begin{align}
h_{{\rm LZ},N}=h_\infty + (a \log_2(N))/N^\gamma, \label{eq:LZcplx_extrapolation}
\end{align}
motivated
in Ref.\ \cite{Schuermann1996} and also used in Ref.\ \cite{Lesne2009}, was
employed. The main plot of Fig.\ \ref{fig:LZcplx}(a) shows the extrapolated
asymptotic entropy rates $h_\infty(T)$. 
The asymptotic entropy rates are in accord with intuition: at low $T$, the 
symbolic sequences exhibit a high degree of order, hence the associated 
entropy rate is small (vanishing in the limit of perfect order). 
In contrast, at high $T$, the sequences exhibit maximal randomness, i.e.\ subsequent spin 
orientations in the underlying model are uncorrelated, and the entropy rate tends towards
$h_\infty=1$. Of pivotal interest is the region close to $T_{\rm c}$ where nontrivial, long-ranged correlations
between the successive orientations of the monitored spin build up. Below it will be of interest
to check whether the previously introduced measures of statistical complexity are sensitive to these
structural changes and can be used to locate the critical point by means of a finite-size scaling analysis
using sequences of different length $N$.
The bottom inset of the figure indicates the change of the fit-parameter $\gamma$ as function
of the temperature. As evident from the figure, in the high-temperature regime
above the critical point $T_{\rm c}=2.269\ldots$, it
assumes a stationary value $\gamma(T>T_{\rm c})\approx 0.73(2)$. In the
low-temperature regime it exhibits an increasing value with decreasing
temperature.  Overall, the agreement between the asymptotic entropy rate
$h_\infty$, the entropy rate $h_{{\rm LZ},N}$ at $N=2^16=65536$ and the common block-entropy at
$h_{{\rm BE},N}$ at $N=10^5$ is remarkably good, see Tab.\ \ref{tab:tab1}. In
particular, for all values of $T$ considered, $h_{{\rm LZ},N}$ seems to be a
satisfactory approximation to $h_\infty$, since both values agree within
errorbars. 

%
\begin{table}[b!]
\caption{\label{tab:tab1}
Comparison of entropy rates for different estimation procedures at different
temperatures. From left to right: System temperature $T$, LZ-parsing based
entropy rate $h_{{\rm LZ},N}$ at $N=2^{16}=65536$, LZ-parsing based asymptotic entropy rate $h_\infty$
(extrapolated using Eq.\ \ref{eq:LZcplx_extrapolation}), and block-entropy
based estimate $h_{{\rm BE},N}$ at $N=10^5$ (block-size 7).  
} 
\begin{ruledtabular}
\begin{tabular}[c]{l@{\quad}lll}
$T$  & $h_{{\rm LZ},N}$ & $h_\infty$ & $h^{{\rm BE},N}$  \\
\hline
2.2   & 0.3442(6) & 0.3441(5) & $0.339(3)$ \\
2.267 & 0.436(1)  & 0.435(1)  & $0.428(3)$ \\
2.5   & 0.647(1) & 0.645(1) & $0.643(2)$ \\
\end{tabular}
\end{ruledtabular}
\end{table}

\paragraph{Excess Entropy:}
Fig.\ \ref{fig:LZcplx}(b) illustrates the block-shuffling based excess entropy
Eq.\ \ref{eq:randShuffleComplexity} as function of the system temperature,
averaged over a large number of input sequences. In the analysis, we restricted
the sum to $M_{\rm max}=23$. As evident from the figure, $C_{\rm s}(T)$ assumes
small values for both, the low-$T$ and high-$T$ regime. In between, i.e.\ in
the paramagnetic phase close to the critical point $T_{\rm c}$, it assumes a
peak value thus satisfying the naive ``one-hump'' criterion
\cite{Crutchfield2000,Feldman1998}. For an increasing sequence length the peak
gets more pronounced and shifts towards $T_{\rm c}$. In order to assess whether
the position of the peak in the asymptotic limit coincides with the critical
point we performed a finite-size scaling analysis. 

To accomplish this, we fitted cubic splines to the peak region  $T\in[2.24,2.5]$ of the excess entropy
data curves to obtain the respective sequence length dependent, thus
``effective'', peak positions $T_{\rm eff}(N)$.  In Fig.\ \ref{fig:LZcplx}(b),
the fit-curves are shown as solid lines (dashed lines are a guide for the eye,
only).  Errorbars for the peak position are computed using bootstrap resampling
of the underlying data \cite{practicalGuide2009}. The asymptotic peak position
is then extrapolated using a fit to 
\begin{align}
T_{\rm eff}(N) = T_\infty + a N^{-b}, \label{eq:asympPeakPos}
\end{align}
yielding $T_\infty=2.269(6)$, $a=O(1)$, and $b=0.35(4)$ (reduced chi-square
$\chi^2_{\rm red}=0.58$), see inset of Fig.\ \ref{fig:LZcplx}(b), in good 
agreement with the literature value $T_{\rm c}\approx 2.269$.  
Albeit not shown here, we further observe that the
fluctuations $N {\rm var}(C_{\rm s})$ are peaked directly at $T_{\rm c}$.

At this point, bear in mind that we study symbolic sequences that represent the
time-series of the orientation of a selected spin, recorded for $N$ MC sweeps
on a $2D$ square lattice of finite side length $L=128$.  Analyzing the specific
heat $C=(k_{\rm B} T^2)^{-1}[\langle E^2 \rangle-\langle E \rangle^2]$
\cite{newman1999} of the $L=128$ $2D$ Ising FM we find an accentuated peak at
$T_{\rm peak}=2.278(1)$, indicating the ``effective'' location of the critical
point for the finite system (not shown). Note that this value is slightly
larger than the asymptotic critical point $T_{\rm c}=2.269\ldots$.  Here, we
obtain the interesting result that, by performing a scaling analysis for the
excess entropy peak-locations for symbolic sequences of different length $N$
(all for the finite system size $L=128$), the results seem to extrapolate
towards $T_{\infty}$ which is in striking agreement with the asymptotic
critical point $T_{\rm c}$.  However, the results are also in reasonable
agreement with the effective critical point suggested by the specific heat.
Hence, within the precision reached by our current analysis we cannot
completely rule out that the results extrapolate towards $T_{\rm peak}$ instead
of the asymptotic critical point $T_{\rm c}$.

\paragraph{Multi-information:}
The results for the standard random shuffle based multi-information Eq.\
\ref{eq:randShuffleMultiInf} are shown in Fig.\ \ref{fig:LZcplx}(c).

Here, a finite-size scaling analysis of the
effective peak positions, again obtained by fitting cubic splines to the 
peak region $T\in[2.24,2.5]$ of the data curves, using bootstrap resampling 
to compute errorbars and Eq.\ \ref{eq:asympPeakPos} to extrapolate to the 
asymptotic limit, yields the estimates $T_\infty=2.268(5)$,
$a=O(1)$ and $b=0.34(2)$ (reduced chi-square $\chi^2_{\rm red}=0.16$).  Similar
to the findings for the excess entropy, the estimate of $T_\infty$ is in good
agreement with the known value of $T_{\rm c}$, indicating that $I_{\rm s}$ is
highly sensitive to the correlations that emerge close to the critical point.

\begin{figure}[t!]
\begin{center}
\includegraphics[width=1.0\linewidth]{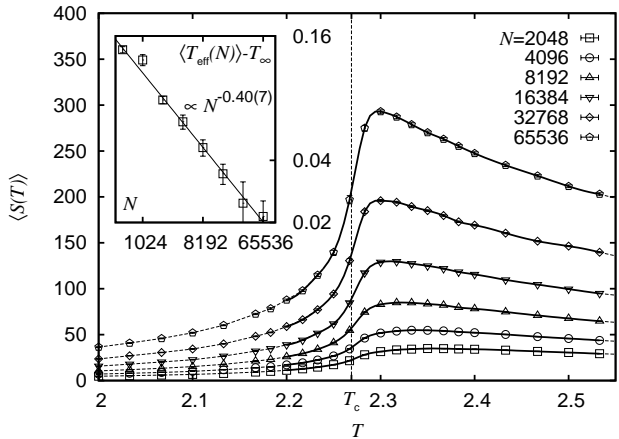}
\end{center}
\caption{\label{fig:Smeasure}
Results for the $\mathcal{S}$-measure as discussed in the text, averaged over many 
input sequences.
The main plot illustrates the $\mathcal{S}$-measure for the LZ parsing 
based entropy rate, considering a standard random shuffle to obtain the 
surrogate sequences used to compute Eq.\ \ref{eq:Smeasure}. The solid lines
represent cubic splines, fitted to the interval $T\in[2.2,2.55]$ and the dashed lines are a guide to eyes, only. The inset
shows the finite-size scaling analysis performed to extrapolate the asymptotic
peak location $T_\infty$.}
\end{figure}

\paragraph{$S$-measure:}
For the purpose of comparing an observable $M$ computed for symbolic sequences of finite length 
to their surrogate counterparts, Ref.\ \cite{JimenezMontano2002b} employed
the $\mathcal{S}$-measure
\begin{align}
\mathcal{S}[S]=\frac{|M_{\rm orig}[S]-\langle M_{\rm surr}[S] \rangle|}{{\rm sDev}(M_{\rm surr}[S])}. \label{eq:Smeasure}
\end{align}
Therein, in order to quantify a significant deviation between both observables,
it states the difference between the observable for the original sequence to
the average value of the observable for an ensemble of proper surrogates,
measured in units of the standard deviation found for the surrogate ensemble.
Here, to probe the sensitivity of the $\mathcal{S}$-measure to structural
changes in the symbolic sequences at different temperatures, we choose as an
observable the LZ complexity based estimator for the entropy rate. I.e.\ for a
given sequence $S$ we consider $M_{\rm orig}[S]=h_{\rm LZ}[S]$ and $M_{\rm
surr}[S]=h_{\rm LZ}[S^{(1)}]$ (note that this corresponds to the construction procedure 1
for surrogate sequences in Ref.\ \cite{JimenezMontano2002b}), using $10^2$
surrogate sequences for averaging obtained by standard random shuffling.
In Eq.\ \ref{eq:Smeasure}, the average $\langle \cdot \rangle$ and standard deviation 
${\rm sDev}(\cdot)$ are computed from $100$ independent surrogate sequences.
Fig.\ \ref{fig:Smeasure} illustrates the $\mathcal{S}$-measure, averaged over
different (original) sequences for various values of $N$. 
As evident from the figure 
the $\mathcal{S}$-measure exhibits an isolated peak close to the critical
point.  This does not come as a surprise: albeit normalized by a temperature
dependent quantity, the enumerator effectively matches the random shuffle based
multi-information Eq.\ \ref{eq:randShuffleMultiInf}.
Here, a finite-size scaling analysis of the system-size dependent peak locations
(obtained by fitting cubic splines to the data points in the interval $T\in[2.2,2.55]$) yields $T_\infty=2.277(10)$, $a=O(1)$, and
$b=0.40(7)$ (reduced chi-square $\chi^2_{\rm red}=1.11$), in agreement with the above results.  

\subsubsection{Common data compression utilities}

In the previous subsection we analyzed three different observables that appear
to be very sensitive to structural changes in the symbolic sequences of finite
length, as the critical point of the underlying model is approached. These were
the standard random shuffle based excess entropy (also termed ``complexity''),
multi-information and the $\mathcal{S}$-measure for the LZ parsing based
entropy rate.  Subsequently, we will restrict our further analysis to the
shuffling based multi-information since it is very simple to compute 
and seems to be able to detect and quantify structural changes in the recorded sequences
that might be used to locate the phase transition point of the underlying model with ease.
Furthermore it has a clear cut interpretation: it yields the entropy rate difference observed
for a symbolic sequence including all symbol correlations and a surrogate sequence featuring the
same symbol frequencies without correlations.

%
\begin{table}[b!]
\caption{\label{tab:tab2}
List of the asymptotic critical points $T_{\rm c}$ and scaling exponents $b$,
estimated from the finite-size scaling (see Eq.\ \ref{eq:asympPeakPos}) of the
peak locations associated to the standard random shuffle based
multi-information, implemented by means of the {\rm zlib}, {\rm bz2}, and 
{\rm lzma} data compression utilities. Therein, the peaks where fit
 by cubic splines (CS) and polynomials 
of order 8 (P8). To facilitate comparison, note that 
the known critical temperature of the $2D$ Ising FM reads $T_{\rm c}=2.269\ldots$.
} 
\begin{ruledtabular}
\begin{tabular}[c]{l@{\quad}llll}
 & \multicolumn{2}{c}{CS} & \multicolumn{2}{c}{P8}\\
\cline{2-3} \cline{4-5}
        & $T_{\infty}$ & $b$ & $T_{\infty}$ & $b$   \\
\hline
ZLIB    & 2.278(5) & 0.5(1)  & 2.24(1)  & 0.24(2)  \\ 
BZ2     & 2.26(1)  & 0.28(7) & 2.265(4) & 0.31(12) \\
LZMA    & 2.276(4) & 0.41(5) & 2.274(9) & 0.39(7)  \\
\end{tabular}
\end{ruledtabular}
\end{table}

As pointed out above, we here consider three commonly used black-box data compression 
tools, namely {\rm zlib} \cite{zlib_ref}, {\rm bz2} \cite{bzip2_ref}, and {\rm lzma} \cite{lzma_ref}, 
in order to compute an algorithmic entropy that is based on the compressibility of the
underlying sequence according to Eq.\ \ref{eq:algEntropy}, see Ref.\ \cite{Ebeling1997}.

\begin{figure}[t!]
\begin{center}
\includegraphics[width=0.9\linewidth]{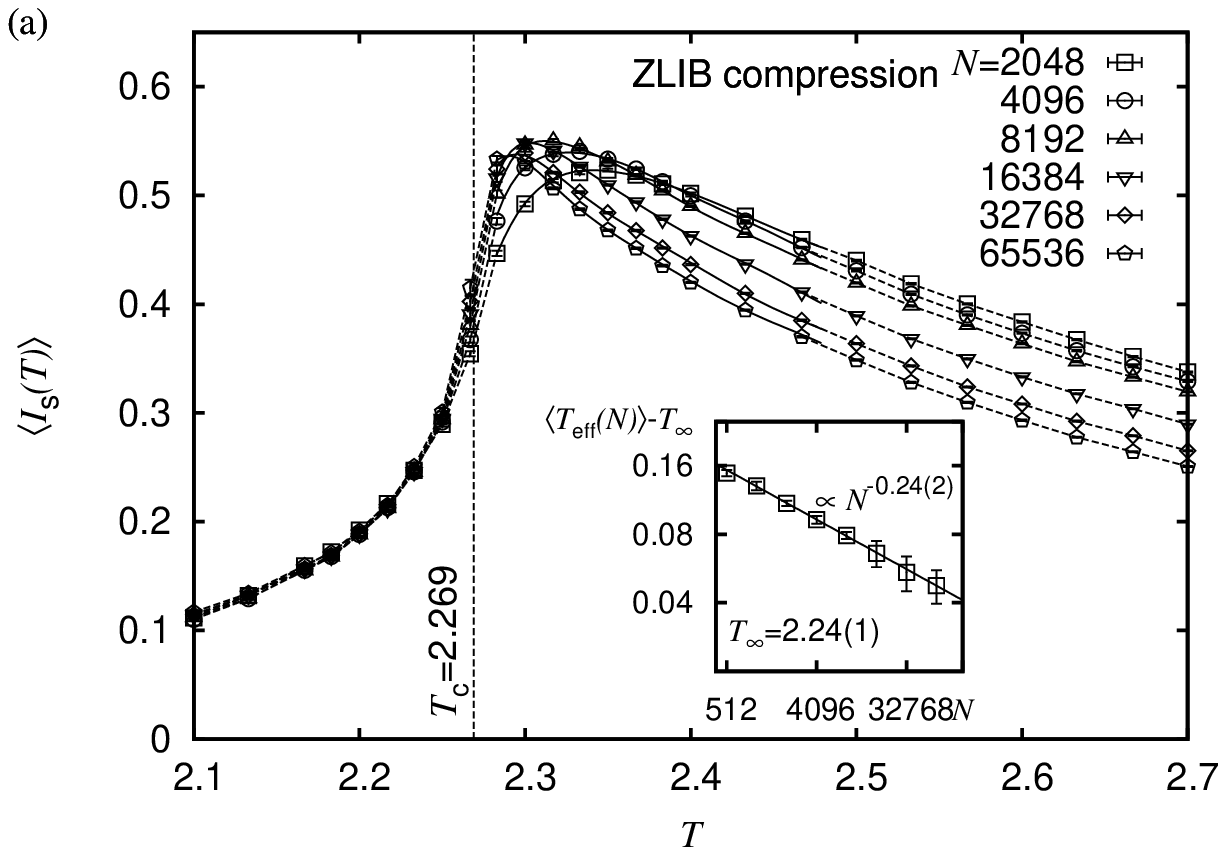}
\includegraphics[width=0.9\linewidth]{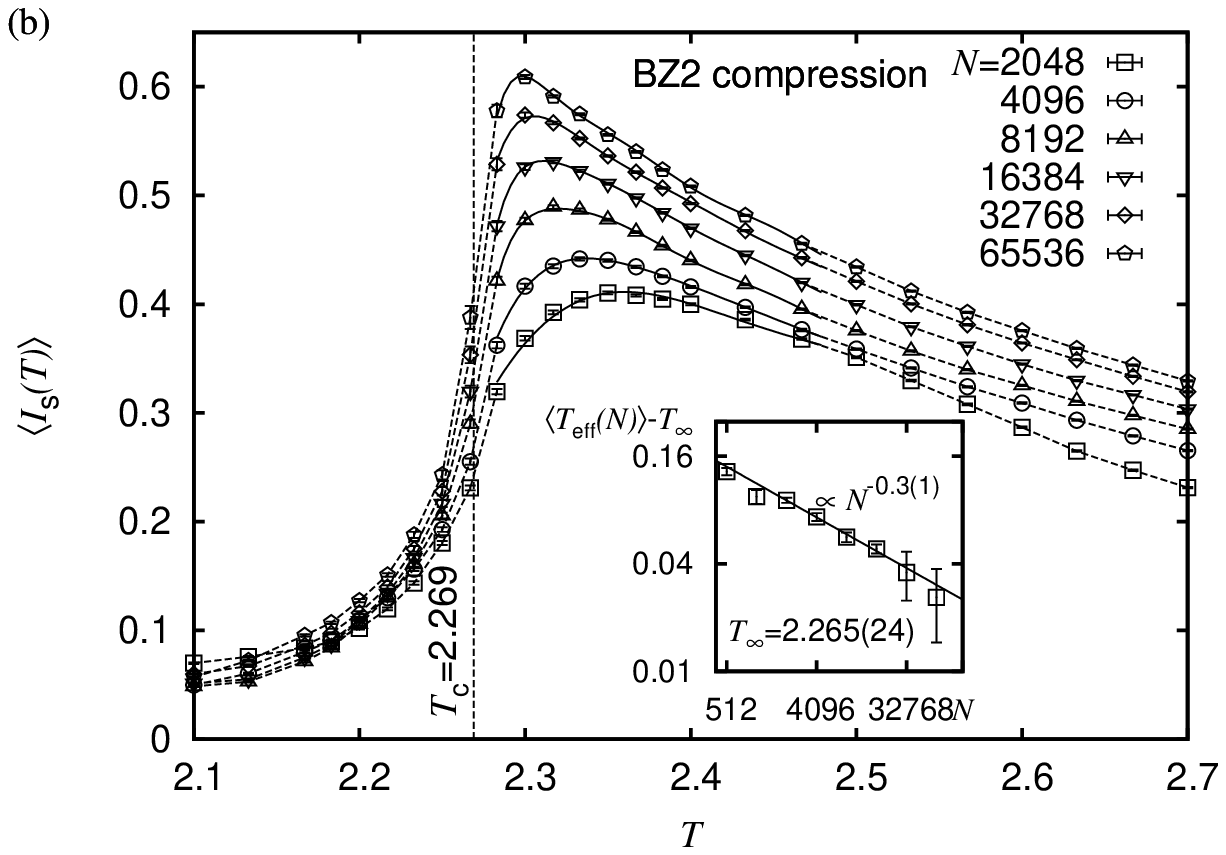}
\includegraphics[width=0.9\linewidth]{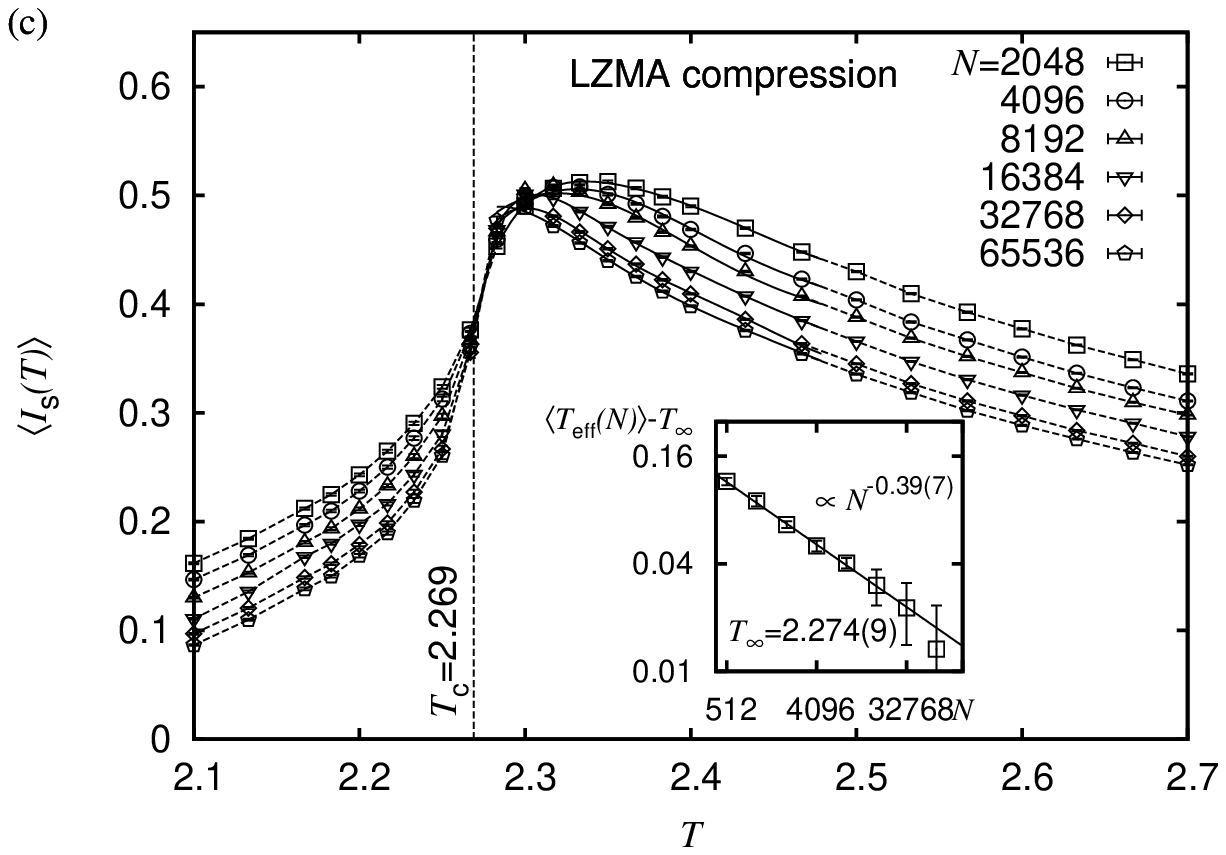}
\end{center}
\caption{\label{fig:multiInfDataCompression}
Results for the standard random shuffle based multi-information computed using three
black-box data compression utilities, averaged over many input sequences. 
The subfigures illustrate the analysis of the shuffling based multi-information using the algorithmic entropy, where 
the ${\rm compress}(\cdot)$ part is implemented using
(a) the {\rm zlib} data-compression tool,
(b) the {\rm bz2} data-compression tool,
(c) the {\rm lzma} data-compression tool.
In the main figures, the solid lines indicate fits to polynomials of order 8, fitted to the interval $T\in[2.24,2.45]$ and the dashed lines are a guide to eyes, only. The insets show the extrapolation of the system size dependent peak locations to the asymptotic limit using
Eq.\ \ref{eq:asympPeakPos}.}
\end{figure}

\paragraph{Results obtained using ${\rm zlib}$:}
The results obtained by implementing the ${\rm compress}(\cdot)$ statement in
Eq.\ \ref{eq:algEntropy} by using the ${\rm zlib}$ data compression tool \cite{zlib_ref} is
shown in Fig.\ \ref{fig:multiInfDataCompression}(a).  
Therein, by fitting the peaks using cubic splines and extrapolating to the
asymptotic limit via Eq.\ \ref{eq:asympPeakPos} we yield $T_\infty=2.278(5)$,
$a=O(1)$, and $b=0.5(1)$ (reduced chi-square $\chi^2_{\rm red}=1.31$), slightly
overestimating the literature value of the critical point $T_{\rm c}$ but still
within a distance of $2\sigma$.  For comparison, the results obtained by
fitting polynomials of order 8 to the data curves are listed in Tab.\
\ref{tab:tab2}.  Regarding the characteristics of the data curves, note that
albeit the peak location monotonously decreases towards a value in decent
agreement with $T_{\rm c}$, the peak height seems to first increase to a value
$\langle I_{\rm s}(T_{\rm eff})\rangle\approx 0.55$ for $5000 \leq N \leq
15000$. For larger value of $N$, the peak height seems to decrease again.

\paragraph{Results obtained using ${\rm bz2}$:}

Implementing the ${\rm compress}(\cdot)$ statement in
Eq.\ \ref{eq:algEntropy} by using the ${\rm bz2}$ data compression tool \cite{bzip2_ref} is
shown in Fig.\ \ref{fig:multiInfDataCompression}(b).  Therein, by fitting the
peaks using cubic splines and extrapolating to the asymptotic limit 
we yield $T_\infty=2.26(1)$, $a=O(1)$, and $b=0.28(7)$ (reduced chi-square $\chi^2_{\rm red}=0.40$), 
in agreement with the literature value of the critical point $T_{\rm c}$.
Again, for comparison, the results obtained by fitting polynomials of order 8
to the data curves are listed in Tab.\ \ref{tab:tab2}. 
As evident from Fig.\ \ref{fig:multiInfDataCompression}(b), and in contrast to the results
obtained using the ${\rm zlib}$ tools, the peak of data curves behaves similar to the multi-information considered  
in the context of the LZ parsing scheme. I.e., the peak consistently shifts towards $T_{\rm c}$ and the 
peak height also increases with increasing $N$ (however, we made no attempt to quantify the latter).

\paragraph{Results obtained using ${\rm lzma}$:}
Lastly, implementing the ${\rm compress}(\cdot)$ statement in
Eq.\ \ref{eq:algEntropy} by using the ${\rm lzma}$ data compression tool \cite{lzma_ref} is
shown in Fig.\ \ref{fig:multiInfDataCompression}(c).  Therein, by fitting the
peaks using splines and extrapolating to the limit $N\to\infty$ 
yields $T_\infty=2.276(4)$, $a=O(1)$, and $b=0.41(5)$ (reduced chi-square $\chi^2_{\rm red}=0.29$), 
in decent agreement with the literature value of the critical point $T_{\rm c}$. 
As can be seen from Tab.\ \ref{tab:tab2}, the results obtained using a fitting procedure
by means of polynomials of order 8 compare even better to $T_{\rm c}$.
Note that here, the peak gets narrower with increasing sequence length $N$
with the peak location shifting towards a value consistent with $T_{\rm c}$. However, here
the effect already observed for the ${\rm zlib}$ tool, i.e.\ a decreasing peak height for
increasing $N$, is even more pronounced.

\begin{figure}[t!]
\begin{center}
\includegraphics[width=1.0\linewidth]{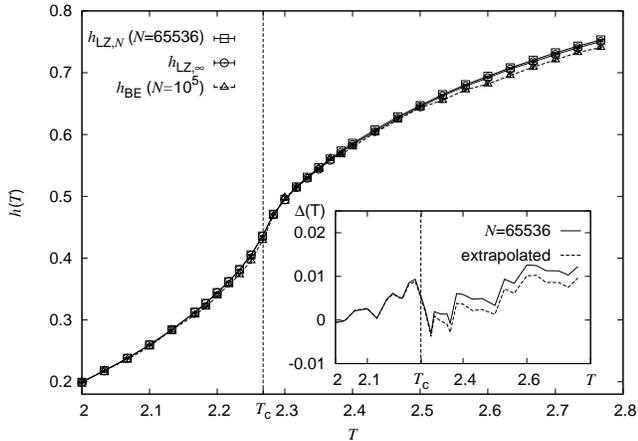}
\end{center}
\caption{\label{fig:LZcplx_eRates}
Comparison of the LZ string parsing scheme based entropy rates to those obtained 
using a well established block-entropy measure.
The main plot contrasts the data curves of $h_{{\rm LZ},N}(T)$ (at $N=65536$), the asymptotic estimate
$h_{{\rm LZ}, \infty}(T)$ and the block-entropy based estimate $h_{{\rm BE},N}(T)$ (at $N=10^5$) as function of the
system temperature $T$. The inset gives an account of the difference 
$\Delta(T)$, referred to in the text.}
\end{figure}

\subsection{Comparison of algorithmic entropy estimates to block entropies}
\label{subsect:compareEntropy}

A further issue we addressed is the question how well the different entropy
estimates (based on the LZ sting parsing scheme and the three data compression
utilities discussed earlier), compare to more conventional block-entropy
estimates, obtained using well established procedures, see Sect.\
\ref{sect:infTheor} and  Refs.\ \cite{Crutchfield2003,Feldman2003}. 

Here, in order to prepare reference values for the entropy rate Eq.\
\ref{eq:entropyRate_difference} using the block-entropy estimator Eq.\
\ref{eq:blockEntropy} we considered sequences of length $N=10^5$ and a maximally
feasible block size $M_{\rm max}=7$, i.e.\ $h_{\rm BE}\equiv h_{\rm
BE}[7]=H_{\rm BE}[7]-H_{\rm BE}[6]$.  As noted earlier, for iid sequences one
might experience a severe undersampling of $M$-blocks if, at a given alphabet
size, $M$ is too large or $N$ is too short. In particular, a naive upper bound
$M_{\rm max}$ might be obtained from the constraint $N\geq M 2^M$
\cite{Lesne2009,Schuermann1996}.  Here, using $M_{\rm max}=7$ and $N=10^5$ we
checked that $h_{\rm BE}[7]$ has converged properly for all considered
temperatures.  Albeit this provides only an upper bound on the true entropy
rate, it might nevertheless yield a reasonable approximation to the actual
entropy of the considered sequences. 

\subsubsection{Lempel-Ziv string parsing scheme}

At first we compared the LZ string parsing based entropy rate estimates to
those obtained using the block entropy estimator. Thereby we performed a
comparison to the entropy rates observed for sequences of finite length
$N=65536$ and to the asymptotic estimates $h_\infty$, resulting from 
extrapolation via Eq.\ \ref{eq:LZcplx_extrapolation}.
The results are illustrated in Fig.\ \ref{fig:LZcplx_eRates}.
Therein, the main plot shows the entropy rate $h_{\rm LZ}(T)$ and $h_{\rm BE}(T)$ 
as function of the system temperature $T$ for the different estimators. The inset indicates
the difference
\begin{align}
\Delta(T) = h_{\rm LZ}(T)-h_{\rm BE}(T)
\end{align} 
between the respective string parsing based entropy rate to the block entropy estimates. 
As evident from the figure, the absolute difference is typically 
smaller than $0.01$ with the largest deviation close to the critical point $T_{\rm c}$.
While both estimates compare similarly well to the block-entropy estimate at low 
temperatures $T<T_{\rm c}$, the extrapolated result $h_\infty$ is slightly closer
to $h_{\rm BE}$ for $T>T_{\rm c}$.

\subsubsection{Common data compression utilities}

\begin{figure}[t!]
\begin{center}
\includegraphics[width=1.0\linewidth]{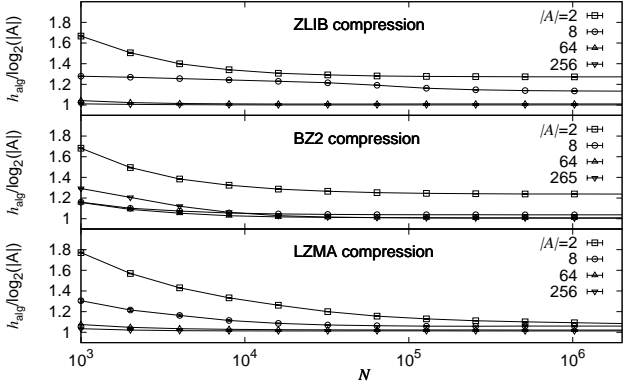}
\end{center}
\caption{\label{fig:compress_eRates_iid}
Comparison of the compression-based entropy rate estimates for iid sequences considering 
the three different data compression tools and alphabet sizes $|\mathcal{A}|=2, 8, 64, 256$.
For increasing alphabet size, the estimate of the entropy rate for a given sequence length 
approach the estimate $\log_2(|A|)$.}
\end{figure}

As reported in the previous paragraph and illustrated in Fig.\
\ref{fig:LZcplx_eRates}, the LZ string parsing based entropy  rate compares
astonishingly well to those estimates obtained using the block-entropy estimator
(even for sequences of finite length $N=65536$). Now, by considering the 
algorithmic entropy rate estimator Eq.\ \ref{eq:algEntropy}, implemented using 
the three data compression utilities discussed earlier, we find that the 
entropy rate estimates for binary sequences of length $N=10^5$ significantly 
overestimate the results obtained via the block-entropy estimator, see the 
data curves for $m=1$, i.e.\ $|\mathcal{A}|=2$ (see discussion below), in Figs.\ \ref{fig:compress_eRates_iid}(a-c).
As mentioned earlier and pointed out in Ref.\ \cite{Ebeling1997}, the
algorithmic entropy rate Eq.\ \ref{eq:algEntropy} provides only an upper bound
on the true entropy rate of the underlying process.  In order to explore
possible routes that might support a more reliable entropy estimate we
next study auxiliary sequences of length $N$, consisting of independent
and identically distributed (iid) symbols, taken from a more general alphabet of size
$|\mathcal{A}|$ instead of binary sequences only.

In Fig.\ \ref{fig:compress_eRates_iid}, we compare the entropy rates obtained
via Eq.\ \ref{eq:algEntropy} using the three data compression tools {\rm zlib},
{\rm bz2}, and {\rm lzma}, to the value $\log_2(|\mathcal{A}|)$ expected for
such iid sequences. As evident from the figure, the respective ratio approaches
unity for increasing alphabet size and for $N$ not too small.  E.g., for
$|\mathcal{A}|=2$ and for large $N>10^5$, the respective estimates read $h_{\rm
alg}/\log_2(2)\approx 1.27$ (zlib), $h_{\rm alg}/\log_2(2)\approx 1.24$ (bz2), $h_{\rm
alg}/\log_2(2)\approx 1.07$ (lzma).  For the larger alphabet-size $|\mathcal{A}|=256$
the respective estimates read $h_{\rm alg}/\log_2(256)\approx 1.0003$ (zlib), $h_{\rm
alg}/\log_2(256)\approx 1.004$ (bz2), $h_{\rm alg}/\log_2(256)\approx 1.07$ (lzma).  Albeit
these results are valid only for iid sequences we might nevertheless expect to
find a tighter upper bound on the entropy rate for symbolic sequences with
possibly long ranged correlations by simply increasing the alphabet-size
$|\mathcal{A}|$.

\begin{figure}[t!]
\begin{center}
\includegraphics[width=0.95\linewidth]{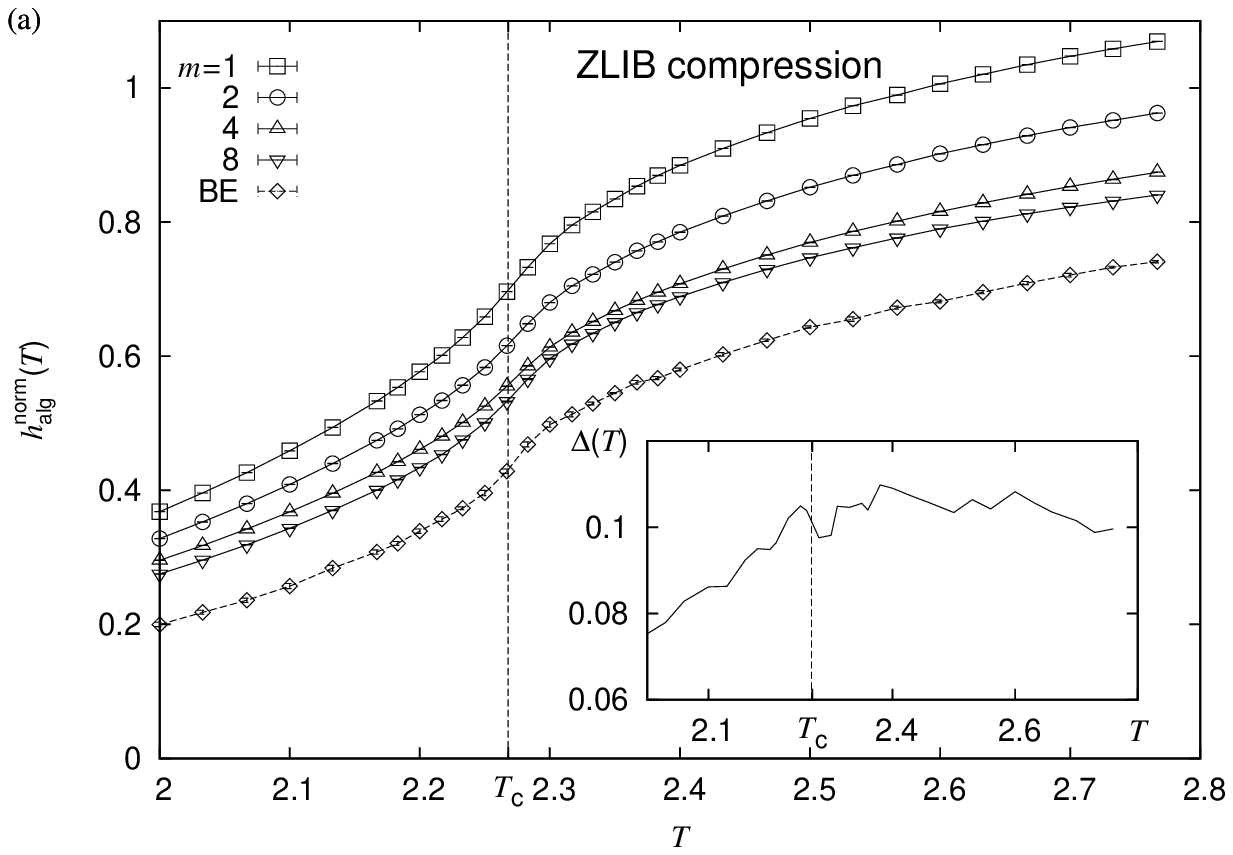}
\includegraphics[width=0.95\linewidth]{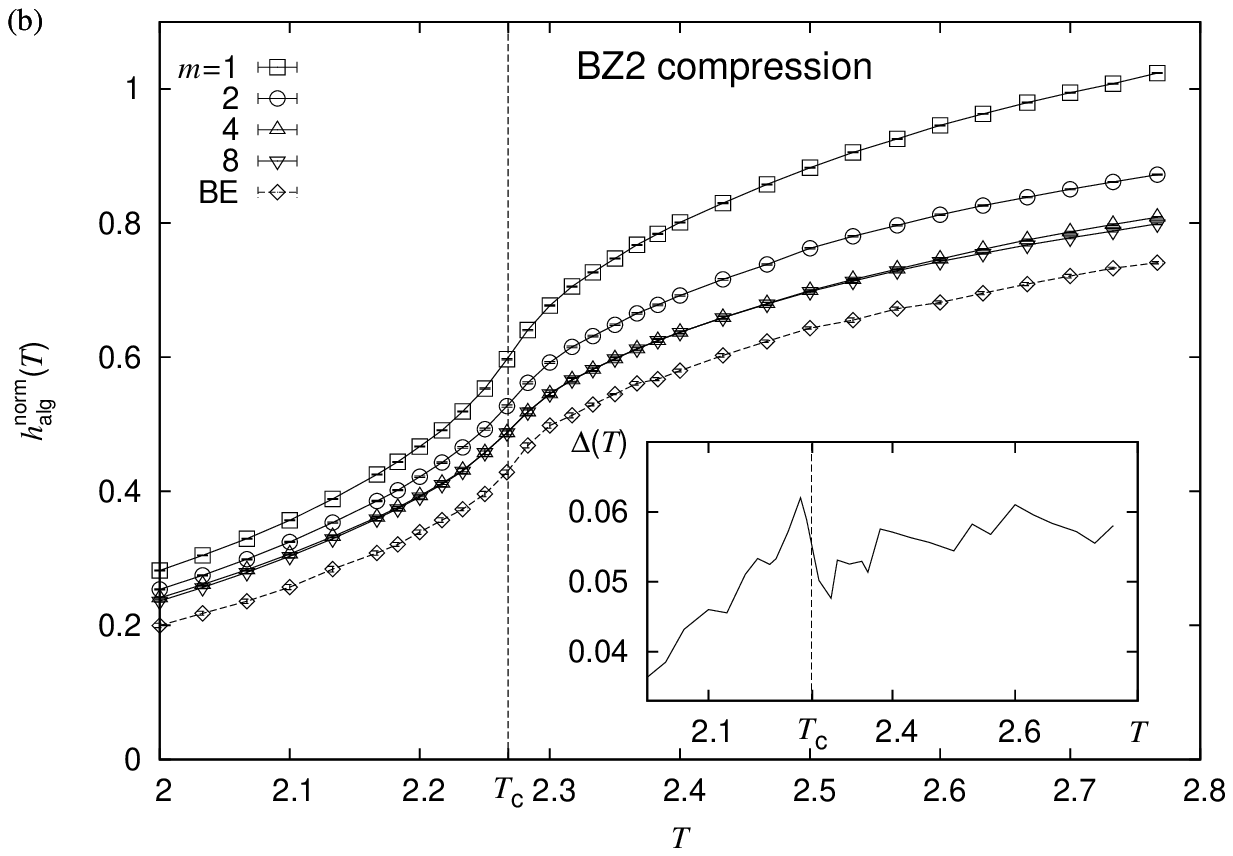}
\includegraphics[width=0.95\linewidth]{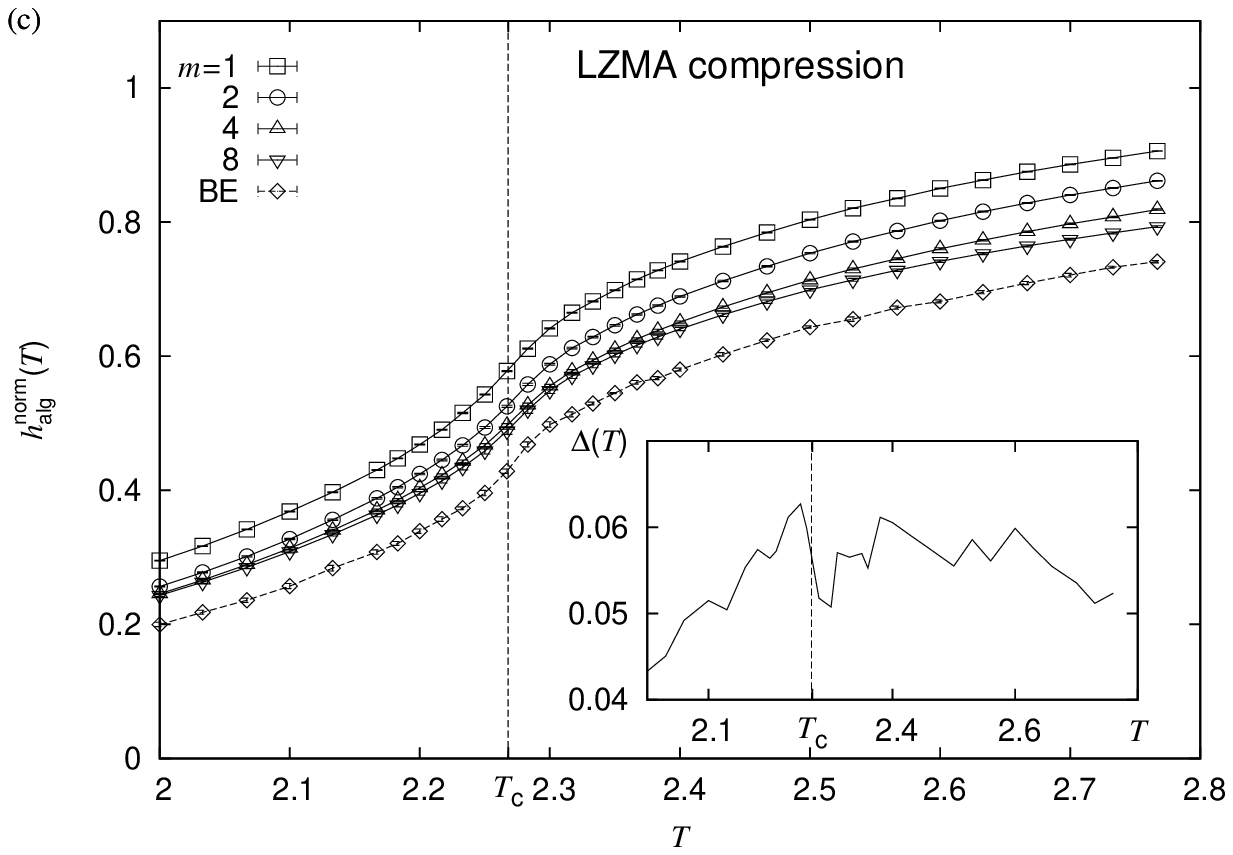}
\end{center}
\caption{\label{fig:compress_eRates}
Comparison of the data compression based entropy rate measures to those obtained 
using a well established block-entropy measure.
The main plot contrasts the data curves of the normalized entropy estimate $h_{\rm alg}^{\rm norm}(T)$ (at $N=10^5$), 
to the block-entropy based estimate $h_{\rm BE}(T)$ as function of the
system temperature $T$ and different neighborhood-template sizes $M$, as discussed in the text. The inset gives an account of the difference 
$\Delta(T)$, referred to in the text.}
\end{figure}

This can be achieved in the following manner: instead to monitor the
orientation of a single spin during the simulation of the $2D$ Ising FM, we
monitor the orientation of a number of, say, $m$ spins, located within a
neighbor-template of size $m$ as introduced in Ref.\ \cite{Robinson2011} (to
analyze the local entropy in a frustrated $2D$ spin system) and used in Ref.\
\cite{Melchert2013} to systematically parse a $2D$ configuration of spins into
$1D$ sequences of length $m$. These can then be interpreted as a binary
representation of a particular symbol from an alphabet of size
$|\mathcal{A}|=2^m$. Following this approach, we show in Figs.\
\ref{fig:LZcplx_eRates}(a-c) the estimates for the entropy rates obtained by
considering neighborhood templates of size $m=1,2,4,8$, i.e.\ alphabet-sizes
$|\mathcal{A}|=2,4,16,256$, to construct symbolic sequences of length $N=10^5$. So as to be able to compare these values to those obtained 
by means of the block-entropy based estimates (monitoring the orientation of a single spin), we normalize the plain algorithmic
entropy estimates resulting from Eq.\ \ref{eq:algEntropy} using $\log_2(|\mathcal{A}|)$. 
I.e.\ we compute 
\begin{align}
h_{\rm alg}^{\rm norm}[S] = \frac{h_{\rm alg}[S]}{\log_2(|\mathcal{A}|)} \equiv \frac{{\rm length}({\rm compress}(S))}{{\rm length}({\rm compress}(S_{\rm iid}))}, \label{eq:algEntropyNorm}
\end{align}
where $S_{\rm iid}$ signifies a iid sequence with the same length and alphabet-size as $S$.
The insets illustrates the difference between $h_{\rm alg}^{\rm norm}$ and $h_{\rm BE}$ for an alphabet-size
$|\mathcal{A}|=256$ as function of the system temperature.
As evident from the Figs.\ \ref{fig:LZcplx_eRates}(a-c), the estimates for increasing alphabet size
indeed approach the block-entropy based estimates. Thereby, the difference $\Delta(T)$ seems to be smallest at 
low temperatures and increases towards higher temperatures, fluctuating around a plateau value for $T>T_{\rm c}$.
Overall, the {\rm lzma} estimator seems to perform best, exhibiting $\Delta\approx 0.05-0.06$ for $T>T_{\rm c}$, 
while the {\rm zlib} based estimator performs worst, exhibiting $\Delta\approx 0.09-0.11$ for $T>T_{\rm c}$.
For comparison, close to the critical point, i.e.\ at $T=2.267$, we find 
$h_{\rm BE}=0.461(7)$, 
$h_{\rm alg}^{\rm norm}=0.532(1)$ ({\rm zlib}),
$h_{\rm alg}^{\rm norm}=0.487(1)$ ({\rm bz2}),
$h_{\rm alg}^{\rm norm}=0.4884(6)$ ({\rm lzma}).


\section{Summary}
\label{sect:summary}
In the presented article we considered information theoretical observables to
analyze short symbolic sequences, comprising binary time-series that represent
the orientation of a single spin in the $2D$ Ising ferromagnet, for different
system temperatures $T$. The latter were chosen from the interval $T\in
[2,2.8]$, enclosing the critical point $T_{\rm c}\approx 2.269$ of the model.
Here, our focus was set on the estimation of the entropy rate via (i) a
Lempel-Ziv based string-parsing scheme, and, (ii) common data compression
utilities (in particular: {\rm zlib} \cite{zlib_ref}, {\rm bz2}
\cite{bzip2_ref}, and {\rm lzma} \cite{lzma_ref}).
These approaches requiere a much smaller computational effort
compared to the standard block entropy approach. Furthermore, they
can be considered as simple yet useful versions of ``algorithmic''
entropy calculations which in principal seek the shortest of all 
programs generating a given sequence.

In a first analysis we demonstrated that certain standard random shuffle based
variants of the excess entropy, multi information as well as an entropy-rate
related $\mathcal{S}$-measure might be used to obtain reasonable estimates for
the critical point of the underlying model.  Albeit we obtained good results
for all three observables when considering the LZ string parsing scheme, we
restricted our analysis of the common data-compression tools to the multi
information $I_{\rm s}$ since it was easy to compute by means of black-box data compression
utilities. As evident from Tab.\ \ref{tab:tab2}, the estimated critical
temperatures, obtained by an extrapolation of the multi-information
peak-location via Eq.\ \ref{eq:asympPeakPos}, compare well to the known critical
temperature.
As pointed out earlier, we here obtain the interesting result that, by
performing a scaling analysis for the multi-information peak-locations for
symbolic sequences of different length $N$ (all for the finite system size
$L=128$), the results seem to extrapolate towards values $T_{\infty}$ (see Tab.\ \ref{tab:tab2}) which are
in striking agreement with the asymptotic critical point $T_{\rm c}$.  However,
the results are also in reasonable agreement with the effective critical point
$T_{\rm peak}$, indicated by the accentuated peak of the specific heat for the
$L=128$ square lattice.  Hence, within the precision reached by our current
analysis we cannot completely rule out that the results extrapolate towards
$T_{\rm peak}$ instead of the asymptotic critical point $T_{\rm c}$.
Note that an unrelated study Ref.\ \cite{Vogel2012}, where 
a data-compression tool for the recoginition of magnetic phases was designed
(based on a different algorithmic procedure and using different observables to locate
the critical point), 
found $T_{\rm C}\approx 2.29$ (for $L=128$) and $T_{\rm C}\approx 2.28$ (for $L=256$) and loosely 
concludes that the findings extrapolate to the known asymptotic critical point. 
Also note that conceptually similar analyses, considering block-entropy based
observables carried out on two-dimensional configurations of spins obtained
from a simulation of the $2D$ Ising FM, reported in Ref.\ \cite{Feldman2008},
conclude that the excess entropy is peaked at a temperature $T_c\approx 2.42$
in the paramagnetic phase slightly above the true critical temperature.
Similar results on the mutual information \cite{Crutchfield2003} for the $2D$
Ising FM (and more general classical $2D$ spin models) where recently presented
in Ref.\ \cite{Wilms2011}. Therein, the authors conclude that the mutual
information reaches a maximum in the high-temperature paramagnetic phase close
to the system parameter $K=J/k_BT\approx0.41$ (for $J=k_B=1$ this corresponds
to $T\approx 2.44$). Our new results and analyses,  which go beyond the cited
literature are presented in our main result part Sec.\ \ref{sect:results}.

In a second analysis we first prepared benchmark data curves for the asymptotic
entropy rate of the symbolic sequences via a block-entropy based approach.
Subsequently we compared the results of the various algorithmic entropy
estimators to the latter.  We found that the LZ string parsing scheme yields
entropy rate estimates (either, for finite sequence length and extrapolated to
the asymptotic limit) that compare surprisingly well to the benchmark data
curves.  Further, for the data-compression based estimators we discussed an
approach that allows to increase the size of the alphabet $\mathcal{A}$ from
which symbols are drawn by monitoring a neighborhood-template
\cite{Robinson2011,Melchert2013}, instead of a single spin, only.
This was motivated by the observation that data-compression based 
estimators strongly overestimate the entropy rates used as a benchmark. 
Consequently, symbolic sequences obtained by means of the amended approach
encode temporal as well as spatial correlations between the orientation of the
spins within the chosen neighborhood-template.  We found that for an increasing
alphabet size, the normalized entropy rates for
the sequences approach get closer to the benchmark estimates, supporting the
intuition previously gained by analyzing iid sequences.  However, the observed
difference between both might be due to the finite dictionary size employed
during the data-compression procedures and hence the inability of the
data-compression based estimators to take advantage of long ranged correlations
in the symbolic sequence.  In principle we found that the {\rm lzma} ({\rm
zlib}) based entropy rate estimator performs best (worst).


\begin{acknowledgments}
OM acknowledges financial support from the DFG (\emph{Deutsche Forschungsgemeinschaft})
under grant HA3169/3-1.
The simulations were performed at the HPC Cluster HERO, located at 
the University of Oldenburg (Germany) and funded by the DFG through
its Major Instrumentation Programme (INST 184/108-1 FUGG) and the
Ministry of Science and Culture (MWK) of the Lower Saxony State.
\end{acknowledgments}


\bibliographystyle{unsrt}
\bibliography{lit_entropyComplexity.bib}

\end{document}